# WHY WE NEED AN INDEPENDENT INDEX OF THE WEB

Dirk Lewandowski, Hamburg University of Applied Sciences
dirk.lewandowski@haw-hamburg.de

Search engine indexes function as a "local copy of the web"[1], forming the foundation of every search engine. Search engines need to look for new documents constantly, detect changes made to existing documents, and remove documents from the index when they are no longer available on the web. When one considers that the web comprises many billions of documents which are constantly changing, the challenge search engines face becomes clear. It is impossible to maintain a perfectly complete and current index.[2] The pool of data changes thousands of times each second. No search engine can keep up with this rapid pace of change.[3] The "local copy of the web" can thus be viewed as the holy grail of web indexing at best — in practice, different search engines will always attain a varied degree of success in pursuing this goal.[4]

Search engines do not merely capture the text of the documents they find (as is often falsely assumed). They also generate complex replicas of the documents. These representations include for instance information on the popularity of the document (measured by the number of times it is accessed or how many links to the document exist on the web), information extracted from the documents (for example the name of the author or the date the document was created), and an alternative text-based version of the document comprising the anchor texts from other documents which link to it. It is important to distinguish between simply finding and saving documents and performing the additional work involved in processing and preparing the documents for search engine use.

Another factor to consider is the user's perspective. Everyone uses web search. Using a search engine to look things up on the web is the most popular activity on the internet, ranking higher even than email.[5] Search engines are used for all types of research. Everything from simply finding a previously visited website to trivia and complex queries such as planning a vacation or treatment options for illnesses.

One amazing thing about search is that users rely predominantly on one particular search engine, Google. Google has a near monopoly in European countries, where it commands a market share of over 90%.[6]

Search engine users therefore must not only rely on Google as a search engine and its unique method of ordering results, they are also confined by the limitations of Google's collection of data. If Google hasn't seen it — and indexed it — or kept it up to date, it can't be found with a search query.

---

[1] Risvik, K. M., & Michelsen, R. (2002). Search engines and web dynamics. *Computer Networks*, *39*(3), 289–302.

[2] Risvik & Michelsen (2002); Lewandowski, Dirk: Suchmaschinen. In: *Grundlagen der praktischen Information und Dokumentation.* 6. Ausgabe. Berlin: De Gruyter; 2013:495–508.

[3] Lewandowski, D. (2008). A three-year study on the freshness of Web search engine databases. *Journal of Information Science*, *34*, 817–831; Ntoulas, A., Cho, J., & Olston, C. (2004). What's new on the web?: the evolution of the web from a search engine perspective. In *Proceedings of the 13th international conference on World Wide Web* (pp. 1–12).

[4] Studies show that search engines do not come close to achieving a comprehensive index and that different search engines do not always find the same documents. Although there is no current research on the comprehensiveness of search engines, from older research (Bharat, K., & Broder, A. (1998). A technique for measuring the relative size and overlap of public Web search engines. *Computer Networks and ISDN Systems*, *30*(1-7), 379–388; Lawrence, S., & Giles, C. L. (1998). Searching the world wide web. *Science*, *280*, 98–100; Lawrence, S., & Giles, C. L. (1999). Accessibility of information on the web. *Nature*, *400*(8), 107–109; Gulli, A., & Signorini, A. (2005). The indexable web is more than 11.5 billion pages. In *14th international conference on World Wide Web* (pp. 902–903).) We can however assume that even when taking into consideration recent technological advances, the coverage of the web is likely to be far from complete.

[5] Purcell, K., Brenner, J., & Raine, L. (2012). *Search Engine Use 2012*. Pew Internet & American Life Project. http://pewinternet.org/~/media/Files/Reports/2012/PIP_Search_Engine_Use_2012.pdf ; Van Eimeren, B., & Frees, B. (2012). ARD/ZDF-Onlinestudie 2012: 76 Prozent der Deutschen online – neue Nutzungssituationen durch mobile Endgeräte. *Media Perspektiven*, (7-8), 362–379.

[6] Lunapark. (2013). Suchmaschinen-Marktanteile. Lunapark. Retrieved from http://www.luna-park.de/blog/1175-suchmaschinen-marktanteile/ ; Schmidt, H. (2012, March 12). Googles Marktanteil steigt auf 96 Prozent in Deutschland. *Focus Online*; http://www.focus.de/digital/internet/netzoekonomie-blog/suchmaschinen-googles-marktanteil-steigt-auf-96-prozent-in-deutschland_aid_723240.html

But what about the alternatives? If Google is handling over 90% of search requests, then at least the remaining 10% are going to other search engines. Here it is important to note that many providers of what may appear to be a search engine are simply services which access the data of another search engine, representing nothing more than an alternative user interface to one of the more well-known engines. And in many cases, that turns out to be Google.

The larger internet portals have not run their own search engines for quite some time. Instead, they rely on their services partners. If we take Germany as an example, we can see that the major internet portals T-Online, GMX, AOL, and web.de all display results obtained from Google. Consequently, Google's market share continues to grow. A significant proportion of those who believe they are using an alternative search engine are accessing Google's data whether they know it or not.

True alternatives, which do exist, only play a very minor role in the German search engine market. The main contenders here are Bing, Ask.com, and Metager. But of these, only Bing can claim a significant share of search requests. Another interesting fact is that one of these alternatives is a meta-search engine which does not have its own index. Instead, Metager accesses the databases of several other search engines.

The search engine market in other European countries is similar to Germany's. Google dominates and other search engines can only achieve minimal market share or don't represent a true alternative since they cannot display their own search engine results, only Google's.

One can of course view the decision to use the Google search engine as a personal choice and ask what's so bad about the fact that one search engine is used almost exclusively. The simple answer is that, much in the same way we require more than just a single newspaper to ensure a diverse range of opinions are represented in the media, we need more than one search engine to ensure that a broad range of opinions are represented in the search market.

The comparison with the media becomes somewhat less applicable when one considers that search engines do not play the role of gatekeepers quite the same way as media outlets do when they select a relatively small number of items from a large quantity of existing content. With search, every request generates a new collection of documents from which the user can then make their selection. So what we end up with is a double or, more precisely, a two-stage selection process. The selection made by entering a search request as well as the selection of one or several of the results which are presented.

Diversity in the search engine marketplace would be desirable to ensure that the selection does not always occur according to the same criteria and users have the choice between different worldviews which originate as a product of algorithm-based search result generation.

Which does not mean that the search engines are promoting any specific worldviews, consciously or otherwise. But the decisions required to implement algorithms for ranking documents are influenced by factors which go far behind purely technical considerations. Consequently, an ideology-free ranking algorithm is not possible and would also not be desirable. In its place, we should strive for diversity achieved through multiple ranking algorithms competing against one another.

In the following, I will attempt to illuminate the critical role indexes play in the diversity of the search landscape. I argue the merits of an independent index which is accessible to everyone and show that the lack of such an index can be viewed as a disincentive for investing in search.

**Alternative search engine indexes**
There are only a handful of search engines that operate their own indexes. Particularly among new search engines startups, it can frequently be observed that they prefer to either rely on an existing external index, limit themselves to a very specialized topic (which frequently requires only a very small index), or they aggregate data from a number of different search engines to offer what is known as a meta search engine.

Meta search engines don't directly access the indexes of the search engines they collect their data from. Instead, they receive only a certain number of high-ranking results from each of their source search engines which are then assembled to form a new ranking. Consequently, a meta-search engine that aggregates the results of five other search engines, receiving 20 items each from them, will have a maximum of 100 documents listed in its own ranking. It's easy to see how meta search engines are relatively limited. In addition, they don't have access to the documents themselves. Rather, they only have the URLs that are provided with the descriptions in the results lists of the source search engines together with the ranking information from the source search engines (i.e. how high the respective document ranked).

In addition to meta-search engines and specialized search engines[7], there have been a few actual search engine startups in recent years. Providers such as Blekko and Duck Duck Go operate their own indexes. No information is available, however, on the amount of data they have or how frequently it is updated. And they are more the exception than the rule in any case. Creating and operating a competitive independent web index is simply too expensive (at least for now) to be profitable.[8]

Above I mentioned that some portals and services which appear to be search engines are actually piggybacking on results from "real" search engines. "Real" search engine providers such as Google and Bing operate their own search engines but also provide their search results to partners. Yahoo, for instance, has been displaying results from Bing for years now. On the surface, Yahoo appears to be a search engine (it has its own layout and a somewhat different format for the results than Bing at first glance). The results themselves, however, are identical with those obtained from Bing.

All the major web portals that offer search as just one service among many have now embraced this model. The key component to this type of constellation is what is known as the partner index (the index the search engine provides its partners). Income is earned when the ads accompanying the search results are clicked on. This revenue is split between the search engine provider and its partner. This model is attractive for both sides since the search engine provider encounters only minimal costs in providing the search results to its partners[9], and the operator of the portal no longer needs to go to the great expense of running its own search engine. All that is required is enough traffic on their portal — very little effort is needed to earn a profit with this model. So it's no wonder that there are hardly any alternative search engines around anymore and portals no longer operate their own indexes. The partner index model is simply too lucrative for an alternative solution to be feasible. The partner index model has served to thin out the competition in the search industry.[10] The lack of diversity in the search engine market can be attributed at least in part to the success of this model. After all, the profits from the partner index model are higher the more search result page ads are served. Consequently, large search engines with a comprehensive advertising network have an inherent advantage.[11]

**Access to search engine indexes**
The large search engine indexes can be accessed by means of what are known as application programming interfaces (APIs). This allows partners to automatically obtain search results they can then use for their own purposes. This business model generally allows a certain quantity of free search requests per day, with payment required only after a certain limit is reached. So wouldn't that make it possible to operate a search engine without incurring the cost of creating and maintaining a separate index?

The problem is that even with APIs, there is no direct access to the search engine index. Instead, what is provided is merely a limited number of top results which have already been ranked by the search engine provider. In this way, access via APIs is similar to what is occurring at the meta-search engines (which in some cases also use the APIs) with the difference being that in this case, only one search engine is being used as a source.

---

[7] Specialized search engines can be defined as "those which are limited to a specific subject matter or document property (such as file type)." Lewandowski, D. (2009). Spezialsuchmaschinen. In D. Lewandowski (Ed.), *Handbuch Internet-Suchmaschinen* (pp. 53–69). Heidelberg: AKA, S. 57
[8] It can be assumed that the indexes of the search engines mentioned here are comparably small. Bing, the only real competition for Google, has since it was founded incurred heavy losses which can only be justified by strategic considerations for owning a proprietary web index (2011). Bing macht Milliardenverluste - wie lange wird Microsoft durchhalten? *Password*, (11), 26.)
[9] The largest share of the costs involved in providing a search engine service go towards deploying the search engine (development costs and building and maintaining the index). The costs involved in the processing the individual search requests play only a very minor role.
[10] Lewandowski, D. (2013). Suchmaschinenindices. In D. Lewandowski (Ed.), *Handbuch Internet-Suchmaschinen 3: Suchmaschinen zwischen Technik und Gesellschaft* (pp. 143–161). Berlin: Akademische Verlagsgesellschaft AKA.
[11] It is assumed here that the search results and the ads are obtained from the same provider. This is the most common scenario, but doesn't always have to be the case. Regardless, the search engine that's able to provide the widest range of results from the web would offer the best monetization for the portal operator (assuming the margin for the portal operator remains equal).

For each search result, the API also provides the URL and the description as it is displayed on the results page of the search engine. The document itself, however, is not provided (this would need to be requested separately using the URL). Even more interestingly, the *representation* of the document in the source search engine is also not included. Additionally, the price is determined not only by the number of search requests made, but also by the number of results requested. For example, the Bing API normally provides the top 50 results. Bing charges for additional results beyond the first 50. The total number of results is also limited to 1000, which is frequently insufficient for the purpose of analysis. Other search engines (such as Google) offer only a free version of their API which, however, is limited regarding the search request volume available per day. It's nearly impossible to build a commercial service on this type of limited API.

Indexes that offer complete access do, however, exist. At the top of this list is Common Crawl[12], a nonprofit project that aims to provide a web index for anyone who's interested. The Blekko search engine contributes to this index by donating its own web crawling results to the foundation. There are also a number of web crawling services which make their results available for research purposes, such as the Lemur project[13]. In both cases, the services are merely static crawls which are not updated on a regular basis. Due to this limitation, they are not suitable for use with a commercial search engine. Common Crawl represents an important development. The project has done groundbreaking work in making web indices widely available. The significance of the service should be determined less by whether or not it is actually successful in providing a comprehensive index for the purpose of internet searching and analysis and more by whether or not it succeeds in generating awareness for the topic of publicly available web indexes.

**Alternative search engines**

There have been efforts in the past to establish alternative search engines. But consensus was lacking on what constitutes an "alternative." One might consider all search engines that are not Google as alternatives. This would include all those search engines which have been heralded as the next "Google killer," but never manage to succeed over the long run. The list of failed attempts is long. Cuil certainly counts as one of the higher-profile failures. The launch was preceded by grand claims and great anticipation, but after just a few years, the service was discontinued.

Then there are the search engine alternatives which are not perceived as such because they are considered to be simply the same as Google. This is mainly Bing, but also other providers such as Yahoo, back in the days when it still operated its own search engine. These search engines have in common that, although they are perhaps perceived as being independent providers, they are not seen by users as offering something that goes significantly beyond what Google can provide. They don't offer anything that would make switching search engines worthwhile.

And then there are the search engines which explicitly position themselves as an alternative to Google by emphasizing their superior ability to deal with local or European languages and to control the quality of the search results through their enhanced knowledge of local matters. One such service was the "European search engine" Seekport, which operated between 2004 and 2008.[14] These types of search engines thus do not differentiate themselves by providing a fundamentally different technical approach compared to the dominant search engines. Instead, they focus on a regional approach.

Attempts to provide public support for search engine technology in an effort to promote the creation of one or several alternative search engines should also be noted. One example is the joint German-French project Quaero. After the French and German sides were not able to come to agreement, however, the project was split into two parts, Quaero (the French part) and Theseus (the German part). The French team developed technologies for multimedia searching. Theseus concentrated on semantic technologies for business-to-business applications (without focusing exclusively on search).

"Real alternatives," on the other hand, would be defined as search engines that pursue a different approach than the conventional web search engines. Conceivable services might include semantic search engines and alternative approaches to gathering web content and making it accessible such as web catalogs, social bookmarking services, and collaborative question-answering services. What they all have in common, however, is that their market share is insignificant.

---

[12] See, http://commoncrawl.org/our-work/.
[13] See, http://lemurproject.org/clueweb12/specs.php.
[14] Dominikowski, T. (2013). Zur Geschichte der Websuchmaschinen in Deutschland. In D. Lewandowski (Ed.), *Handbuch Internet-Suchmaschinen 3: Suchmaschinen zwischen Technik und Gesellschaft* (pp. 3–34). Berlin: Akademische Verlagsgesellschaft AKA.

The common thread here is that all of these models and alternatives have failed *as search tools*, even though they may have a role to play as their own genre on the web.

The proposal to provide government funding for search engine technology has been subject to intense criticism in the past. But it's not necessarily a bad idea. This type of support is not appropriate, however, when the objective is to establish only a single alternative. Even if this new alternative were to be of high quality, there would still be a number of factors which would cause it to fail. And they wouldn't even need to have anything to do with search itself. The problem could be poor marketing or the graphic design of the user interface. Regardless of the reason, a failure of the new search engine would result in the entire publicly funded initiative failing.

At this point I would like to revisit the concept of the search engine index. If we were to fund a search engine index and make it available to other providers, it would then be possible to create not just one alternative search engine, but a number of them instead. The failure of one of these search engines would not impact the others. In addition, even if the program is successful, the addition of just a single alternative search engine is not enough — only a critical mass of options (and of course the *actual use* of the new search engines) can truly establish diversity.

If one considers the topic of search engine indexes from an economic perspective, it becomes clear that only the largest internet companies are able to afford large indexes. As I explained above, Microsoft is the only company besides Google to possess a comprehensive search engine index. Yahoo gave up on its own index several years ago, choosing the familiar route of the web portal by sourcing its search results from other providers. It appears as though operating a dedicated index is attractive to practically no one — and there are hardly any candidates with the necessary financial resources in any case. Even if one of the occasional rumors that some company with deep pockets such as Apple preparing to launch its own index were to come true, who would benefit? I am of the opinion that it makes practically no difference whether or not the search engine market is shared by two, three, or four participants. Only by establishing a considerable number of search engines can true diversity be achieved.

**The solution**

The path to greater diversity, as we have seen, cannot be achieved by merely hoping for a new search engine (a Google killer) nor will government support for a single alternative achieve this goal. What is instead required is to create the conditions that will make establishing such a search engine possible in the first place. I have described how building and maintaining a proprietary index is the greatest deterrent to such an undertaking. We must first overcome this obstacle. Doing so will still not solve the problem of the lack of diversity in the search engine marketplace. But it may establish the conditions necessary to achieve that desired end. We cannot predict who will use the data from this index or for what purposes it will be used. But we can expect that the possibilities it presents would benefit a number of different companies, individuals, and institutions. The result will be fair competition to develop the best concepts for using the data provided by the index.

The vision behind this proposal is one that ensures an index of the web can be accessed at fair conditions for everyone.
- "Everyone" means that anyone who is interested can access the index.
- "Fair conditions" does not mean that access to the index must be free of charge for everyone. I assume that the index will be a government-funded initiative provided as infrastructure that should attempt to recoup at least part of the costs involved by charging usage fees. Analogous to the popular API business model described above, a certain number of document requests per day would be available at no cost in order to promote non-profit projects.
- "Access" to the index can be defined as the ability to automatically query the index with ease. Furthermore, it should be easy to actually obtain all the information contained within the index. In contrast to the APIs of the current commercial search engines described above, the complete document representation including the full text of the document should be available. Additionally, there should be no limit with respect to the number of documents or document representations so users of the index can actually create a good ranking based on a large pool of documents. Nevertheless, basic ranking capabilities for querying the index should be provided so that the quantity of documents requested can also be limited in an appropriate manner.[15]

- The concept "index of the web" is intended to cover as much of the web as possible. That means that the index should contain all the common content types currently available from search engines today including news, images, videos, etc. Additional content types that are not currently being indexed by search engines are, however, also a possibility. Bodies of information which pose no interest to consumers but are useful for researchers and developers also exist. Such information includes metadata derived from documents such as frequency distributions of words, information on spam, etc.

This type of project will require a considerable investment of funds. The total cost cannot be precisely forecast here. Several hundred million euros will likely be needed, however, when one considers the anecdotal reports provided by search engine operators. The losses Bing has reported on its search activities are one example. This may appear prohibitively expensive until one considers, for instance, that the German government invested roughly €100 million in the semantic technology developed as part of the Theseus program. Nevertheless, it is still clear that this type of project cannot be supported by any one country alone. The only feasible option is a pan-European initiative.

Not only would the issue of financing need to be settled, the question of who would operate the index must also be considered. An existing research institution or newly-founded institution is one possibility. The institute might operate and maintain the index while also conducting its own research based on the resulting pool of data. The most important thing is that the operator of the index does not obtain the exclusive right to determine the way in which the documents are used or made available. Instead, a board of trustees consisting of (potential) users of the index should participate in the decision-making process regarding document representations and the availability of the data.

In closing, I would like to reiterate the advantages of this type of search engine index. An index of the type described here would motivate companies, institutions, and developers pursuing personal projects to create their own search applications. The data available on the web is so boundless that it lends itself to countless applications in a broad range of fields. A new search engine index would not only make it possible to operate comprehensive web search engines that seek to include practically everything available on the web, it would also enable specialized search engines covering a broad range of subjects. The index would thus be useful not only for entities seeking to operate a general-purpose search engine, but also for libraries, museums, and other institutions seeking to supplement their own data with data from the web. The index would also be useful for applications completely unrelated to search. One example might be analyses based on data from the web for instance to assess public interest in certain topics or the sentiment of users about certain issues over time.

Such an index would also enable applications we are not yet capable of even imagining. An open structure, transparency with respect to access, and the assurance of permanent availability thanks to state sponsorship would lay the groundwork for innovation, even innovation that requires larger investments. An index of this nature will allow anyone to invent new search technology and, more importantly, to test it.

Building and operating a search engine index can thus be seen as part of government's role to provide public infrastructure. The state finances highways used by everyone, ensures that the electrical grid is available to all, and generates and disseminates geodata. Making web data available is no different from these other public services.

---

[15] This should be sufficient in the vast majority of cases. What is important, however, is that the quantity of documents can be determined by users themselves, not by the operator of the index.

6